\documentclass[conference]{IEEEtran}
\ifCLASSINFOpdf
\else
\fi
\usepackage{url}

\usepackage{graphicx}
\usepackage{subcaption}
\usepackage[htt]{hyphenat}
\usepackage[pagebackref=true,bookmarks=false,hidelinks]{hyperref}

\makeatletter
\def\endthebibliography{%
	\def\@noitemerr{\@latex@warning{Empty `thebibliography' environment}}%
	\endlist
}
\makeatother

\begin{document}
%
\title{The Mobile AR Sensor Logger\\ for Android and iOS Devices}

\author{\IEEEauthorblockN{Jianzhu Huai}
\IEEEauthorblockA{Dept. of Civil, Environmental,\\and Geodetic Engineering\\
The Ohio State University\\
Columbus, OH 43210}
\and
\IEEEauthorblockN{Yujia Zhang}
\IEEEauthorblockA{Dept. of Earth and Space Science and Engineering\\
York University\\
Toronto, Ontario, CA}
\and
\IEEEauthorblockN{Alper Yilmaz}
\IEEEauthorblockA{Dept. of Civil, Environmental,\\and Geodetic Engineering\\
The Ohio State University\\
Columbus, OH 43210}}


%


\maketitle

\begin{abstract}
In recent years, commodity mobile devices equipped with cameras and inertial measurement units (IMUs) have 
attracted much research and design effort for augmented reality (AR) and robotics applications.
Based on such sensors, many commercial AR toolkits and public benchmark datasets have been 
made available to accelerate hatching and validating new ideas. 
To lower the difficulty and enhance the flexibility in accessing the rich raw data of typical AR sensors on mobile devices, 
this paper present the mobile AR sensor (MARS) logger for two of the most popular mobile operating systems, Android and iOS. 
The logger highlights the best possible synchronization between the camera and the IMU allowed by a mobile device, 
and efficient saving of images at about 30Hz, 
and recording the metadata relevant to AR applications. 
This logger has been tested on a relatively large spectrum of mobile devices, 
and the collected data has been used for analyzing the sensor characteristics. 
We see that this application will facilitate research and development related to AR and robotics, 
so it has been open sourced at 
\url{https://github.com/OSUPCVLab/mobile-ar-sensor-logger}

\end{abstract}


%
\IEEEpeerreviewmaketitle

\section{Introduction}

"Big data" has been driving the frontier of fields including simultaneous localization and mapping (SLAM) and augmented reality (AR) \cite{goodfellow2016deep}.
A plethora of datasets for benchmarking the SLAM techniques have been made public, e.g.,  \cite{burri2016euroc}\cite{huai2019}. 
While such data offer high integrity and ground truth essential to developing and validating new methods, 
they are often limited in quantity and diversity thanks to the confined labs and expensive devices for data acquisition.

Meanwhile, driven by the consumer market, mobile devices have been decreasing in price and growing in capabilities.
Commodity mobile devices are usually equipped with cameras and inertial measurement units (IMUs) 
which are all it needs for enabling SLAM or AR applications.
Besides the huge user base, many SDKs aiming at SLAM or AR are released and promoted by large corporations, notably, the ARCore and ARKit \cite{linowes2017augmented}.
These facts make it ideal to collect large quantities of data with the mobile devices for research product development related to SLAM and AR at a large scale \cite{huai2017collaborative}. 

Despite the potential, we find it often cumbersome to access the original and complete data from the mobile AR sensors.
Though there are many applications in Google Play or App Store for logging sensor data with diverse flavors, 
few are written with retrieving the full visual and inertial data in mind.
On the other hand, it typically involves a steep learning curve to develop a logging tool meeting such requirements as reasonable synchronization and providing sensor characteristics.
Many times, these tools are used for preliminary investigations and may well be abandoned later on.
To lower the bar of acquiring the full data for mobile AR sensors, 
we propose the MARS logger that runs on two of the most widely used mobile operating systems (OSs), Android and iOS.

\begin{figure}[!tbp]
	\centering
	\subcaptionbox{}{\includegraphics[width=0.40\columnwidth]{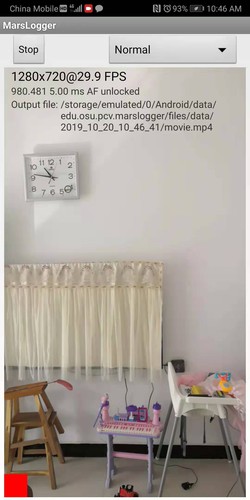}}%
	\hfill
	\subcaptionbox{}{\includegraphics[width=0.45\columnwidth]{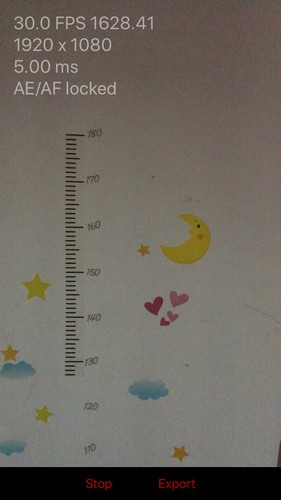}}%
	\caption{MARS logger in use: a, Honor V10 running Android 9; b, iPhone 6S running iOS 12.}
\end{figure}

The MARS logger distinguishes itself with the following features.
1) Image frames are recorded efficiently with the standard APIs (Application Program Interfaces) of mobile OSs. The OpenGL library is used for efficient rendering, while frame encoding and I/O are done in background. In the normal operation mode, frames can be recorded at about 30 Hz over tens of minutes with moderately priced mobile devices. 
2) Rich metadata of frames relevant to SLAM are saved if supported by the devices, including rolling shutter skew and focus distance. In the same spirit, the frames are recorded with auto-focus and auto-exposure locked to suit SLAM methods.
3) Both IMU data and image frames are timestamped and synchronized to one time source.

The next section overviews the related work about data acquisition with mobile devices.
Section~\ref{sec:methodology} discusses the distinctive features of the MARS logger.
Section~\ref{sec:experiments} diagnoses the collected data by calibrating temporal and 
spatial parameters of the camera and the IMU along with motion estimation.
In the end, Section~\ref{sec:conclusion} summarizes the work with remarks.

\section{Related Work}

To tap the potential of AR sensors, large corporations have developed frameworks that
opaquely pass the sensor data to proprietary algorithms which output results for localization and mapping. 
Popular AR frameworks include ARCore for Android and ARKit for iOS \cite{linowes2017augmented}.
Due to the demanding nature and dependency on the hardware,
both frameworks are only available on certain supporting devices.

As the affordable devices proliferate, many SLAM benchmark datasets collected with the mobile sensors have been made available.
Depending on the specific problems, many of them centered about the inertial data for localization  \cite{chen2018oxiod, gjoreski2018university}.
Images from the front cameras were captured infrequently in the UMDAA datasets \cite{mahbub2016active}.
The collaborative SLAM dataset \cite{golodetz2018} provided color and depth frames captured with an Asus ZenFone AR smartphone at 5Hz.
The ADVIO dataset \cite{cortes2018advio} contained sessions lasting up to 7 minutes of camera frames at 60Hz and inertial data at 100Hz captured by an iPhone 6S.  

As the AR field flourishes, numerous applications have been developed for capturing the visual and inertial data.
A few curated ones are reviewed due to the space limit. 
For capturing image frames, the Camera app pre-installed in a device obviously is ideal in terms of efficiency and handiness. 
This app can be paired with an inertial sensor logger running continuously in background for data collection, 
for instance, powersense \cite{xu2017} for iOS, and Sensorstream IMU+GPS \cite{lorenz2013} for Android, with an obvious downside of out-of-sync timestamps. 
There are also the OpenCV SDKs(software development kits) for both Android \cite{opencv2019} and iOS  \cite{opencv2019b}. 
They do not pass timestamps along with images.

As for open sourced applications, the RosyWriter sample from Apple \cite{rosywriter} showed delicate control of the camera and efficient recording of frames. 
The samples in \cite{allan2011} illustrated how to access the raw inertial data in iOS.
For Android, the Google grafika \cite{google2019b} and samples \cite{google2019} showed 
how to use the \texttt{Camera} class and the \texttt{camera2} API for recording videos. 
The RPNG dataset recorder \cite{geneva2016} recorded image sequences and IMU data for Android devices.
The camera recorder \cite{suda2018} captured videos with the Android \texttt{camera2} API.

\section{The Mobile AR Sensor Logger}
\label{sec:methodology}

The MARS logger is designed to acquire visual and inertial data for SLAM and AR applications with commodity mobile devices.
Two versions of the logger is developed to cater to the dominating mobile OSs, Android and iOS.
For better portability and broader coverage, only standard APIs and libraries are employed and deprecated APIs and coding patterns are avoided.
To simplify offline processing by SLAM or AR methods, the focus distance and exposure time of the camera can be locked by a tap on screen before data acquisition.

\subsection{MARS logger in Android}
The MARS logger in Android is developed based on the CameraCaptureActivity of 
the grafika project \cite{google2019b} in Java with the minimum SDK version of API level 21.
Added since API 21, the \texttt{camera2} API returns the camera timestamp and camera intrinsic parameters,
e.g., focal distance and rolling shutter skew, with a \texttt{CaptureCallback}, 
providing richer metadata than the deprecated \texttt{Camera} class.
The sensor timestamp in the capture result is "the time at start of exposure of first row of the image sensor active array, in nanoseconds" \cite{android2019}.
To efficiently display the camera frames on the screen, the \texttt{GLSurfaceView} is used for rendering. 
Additionally, it shares the rendered frame with a \texttt{MediaCodec} encoder,
thus avoiding the pixel alignment issues which arise in using \texttt{ImageReader} and \texttt{MediaCodec} together.
The encoded frames are written to a H.264/MP4 video file by the \texttt{MediaMuxer} sequentially 
and the presentation times for every frames to a text file.

The IMU data are recorded with the \texttt{SensorEventListener} in a background \texttt{HandlerThread}.
To synchronize the accelerometer and the gyroscope, the accelerometer reading is 
linearly interpolated at the epoch of each gyroscope reading.
Each event timestamp is the time in nanoseconds since boot just as returned by \texttt{SystemClock.elapsedRealtimeNanos()}.
If the camera timestamp source is \texttt{CLOCK\_BOOTTIME}, then the camera timestamps are in the same timebase as the IMU data. 
Otherwise, to remedy the incomparable timestamps, the timestamps returned by \texttt{System.nanoTime()} in the timebase of \texttt{CLOCK\_MONOTONIC} 
as well as by \texttt{SystemClock.elapsedRealtimeNanos()} are recorded at the start and the end of a capture session.
In offline processing, the time difference between the two clock sources can be used to correct the camera frame timestamps to the timebase of the IMU data.

\subsection{MARS logger in iOS}
The iOS version is developed 
based on the rosywriter \cite{rosywriter} in Objective C with the iOS SDK 8.0. 
To have more control over camera frames, \texttt{AVCaptureVideoDataOutput} is used together with \texttt{AVAssetWriter} 
to save H.264/MP4 videos instead of \texttt{AVCaptureMovieFileOutput}.
For SDK 11.0 and greater, the \texttt{CameraIntrinsicMatrix} delivery is enabled in the \texttt{AVCaptureConnection} instance for video data,
so that every frame has an attachment containing the focal length and principal point offsets in pixels.
For lower SDK versions, these fields are set to empirical values.
Additionally, the \texttt{exposureDuration} read from the \texttt{AVCaptureDevice} instance is recorded for every frame.

To obtain the IMU data, a \texttt{CMMotionManager} instance is set to update at a specific interval, e.g., 1/100 second, 
and periodically appends arriving \texttt{CMAccelerometerData} and \texttt{CMGyroData} messages to a background \texttt{NSOperationQueue} 
which synchronizes the inertial data similarly to the Android version.
As the inertial data are timestamped by the clock returned by \texttt{CMClockGetHostTimeClock()} 
while the camera frames have presentation times referring to the \texttt{masterClock} of a capture session,
\texttt{CMSyncConvertTime()} is used to convert the frame timestamps to the IMU host time clock for synchronization.

\section{Experiments}
\label{sec:experiments}

Many sessions of visual and inertial data were collected from several handheld devices 
with the MARS logger in order to check the data quality and acquisition performance.

\subsection{Frequency of captured data}

\begin{figure}[]
	\centering
	\subcaptionbox{}{\includegraphics[width=0.45\columnwidth]{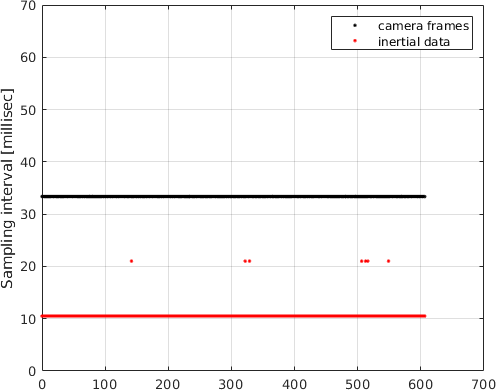}}%
	\hfill
	\subcaptionbox{}{\includegraphics[width=0.45\columnwidth]{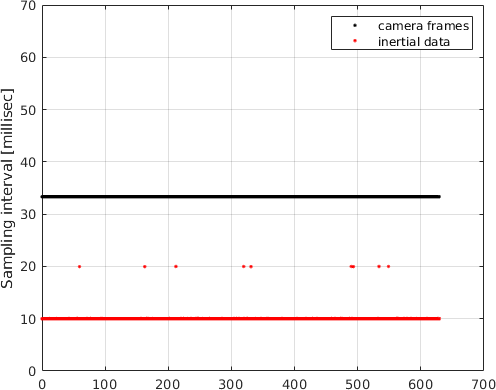}}%
	\hfill
	\subcaptionbox{}{\includegraphics[width=0.45\columnwidth]{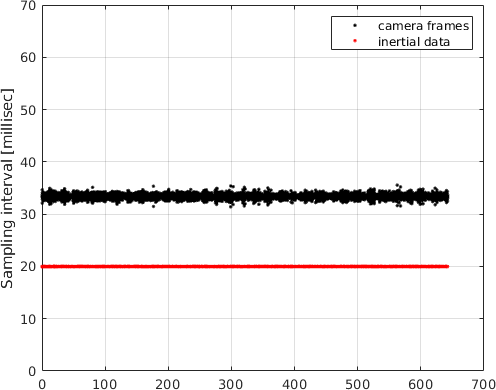}}%
	\hfill
	\subcaptionbox{}{\includegraphics[width=0.45\columnwidth]{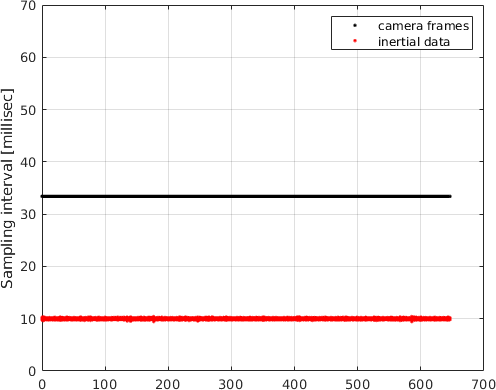}}%
	\hfill
	\subcaptionbox{}{\includegraphics[width=0.45\columnwidth]{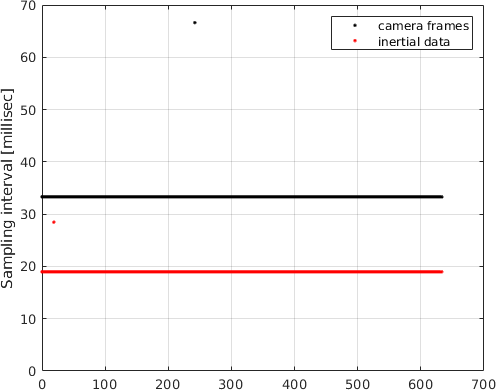}}%
	\caption{Sampling intervals of captured visual inertial data with (a) iPad mini, (b) iPhone 6S, (c) Phab2 Pro, (d) Honor V10, (e) Samsung S9+. 
	Note the inertial data of Android devices were captured at the device-dependent sensor delay \texttt{SENSOR\_DELAY\_GAME}.}
	\label{fig:interval}
\end{figure}

To evaluate the capture efficiency, five devices were used to capture the frames and IMU data,
including an iPhone 6S running iOS 12.3, an iPad mini running iOS 9.3, an Honor View 10 running Android 9.0,
an Lenovo Phab2 Pro running Android 9.0, and an Samsung Galaxy S9+ running Android 8.0.
These devices are valued from \$250 up to \$850, spanning the medium to high price range of mobile devices.
Each device ran the MARS logger for 10 minutes while moving in a nearly static area, 
and the capture intervals of the visual and inertial data were shown in Fig.~\ref{fig:interval}.
The recorded frame size for iPhone 6S was 1920$\times$1080, while it was 1280$\times$720 for other devices.

From the figure, we see that for handheld Android and iOS devices of medium quality, 
the MARS logger can efficiently record camera frames at about 30Hz and IMU data at about 50Hz or 100Hz for over 10 minutes.

\subsection{Sensor parameters and synchronization}

To inspect the quality of the recorded camera intrinsic parameters and timestamps, 
two SLAM algorithms relevant to AR sensor online calibration were chosen for analyzing the datasets.
The VINS Mono \cite{qin2017vins} program were selected to estimate extrinsic parameters and time offset between an camera and an IMU 
for it supports cameras of rolling shutters.
The MSCKF implemented in \cite{huai2017collaborative} were used to estimate the camera intrinsic parameters 
including the rolling shutter skew, i.e., frame readout time, and extrinsic parameters and the time offset between the two sensors.

\begin{figure}[]
	\centering
	\subcaptionbox{}{\includegraphics[width=0.45\columnwidth]{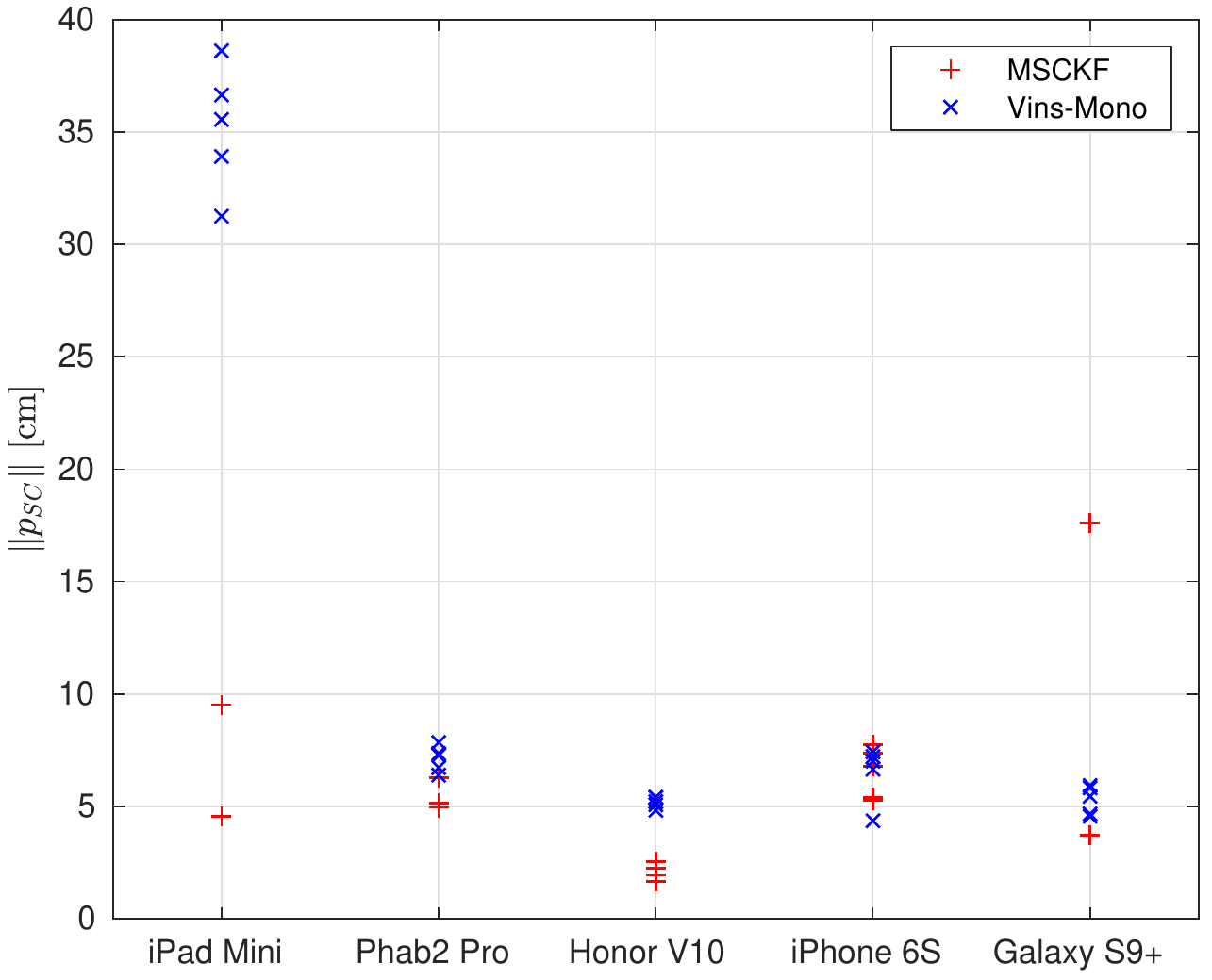}}%
	\hfill
	\subcaptionbox{}{\includegraphics[width=0.45\columnwidth]{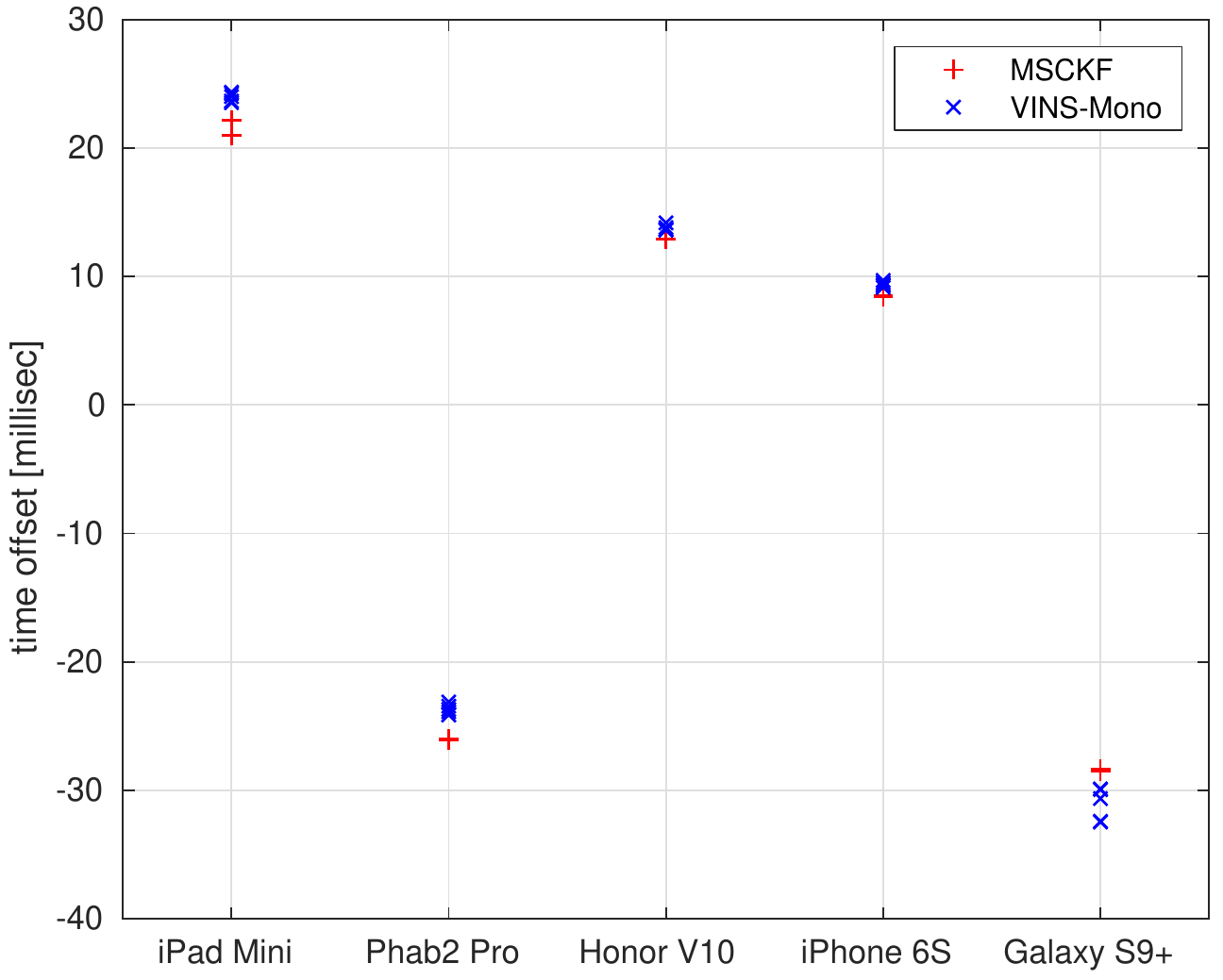}}%
	\hfill
	\subcaptionbox{}{\includegraphics[width=0.45\columnwidth]{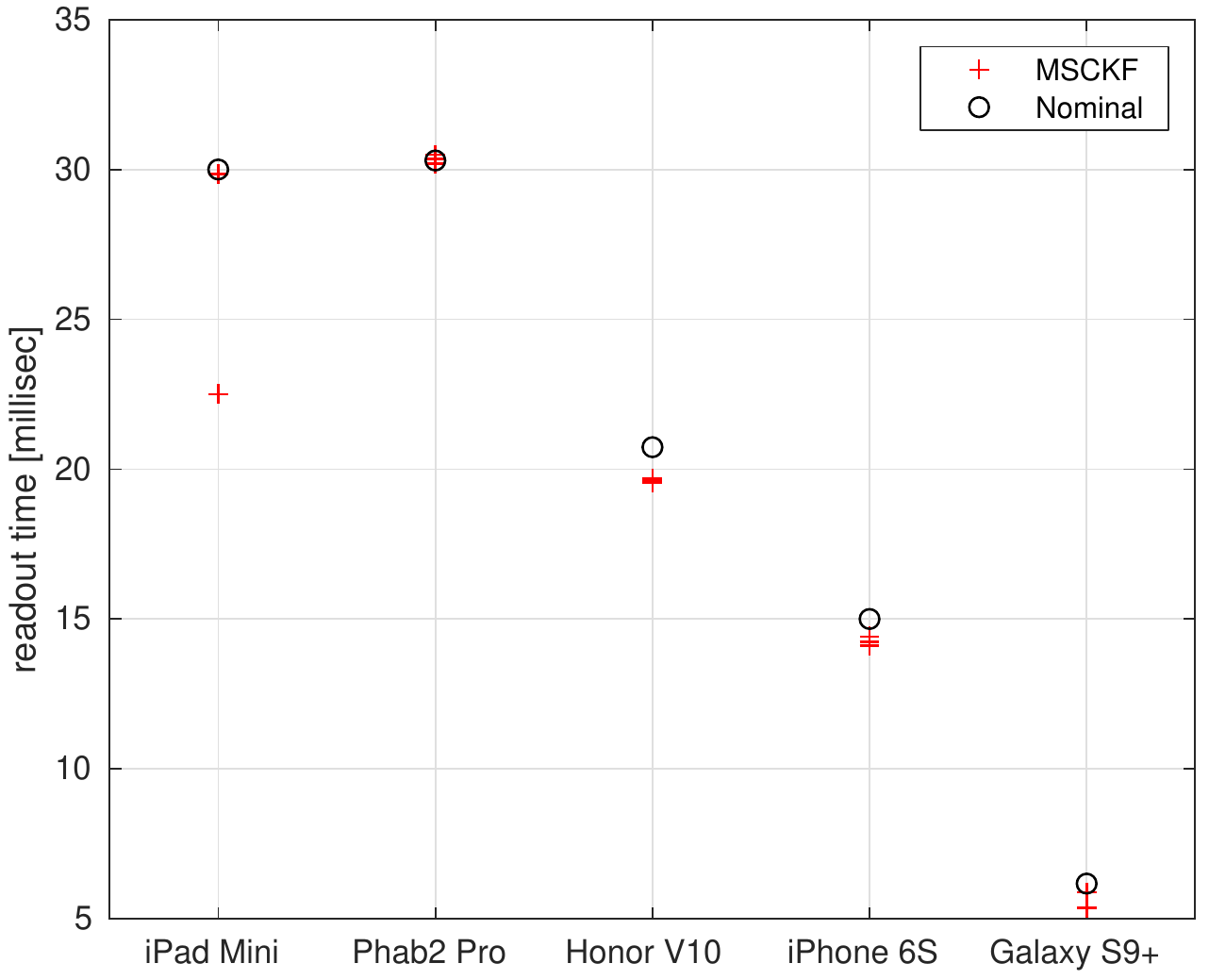}}%
	\caption{The camera-IMU lever arm (a), time offset (b), and frame readout time (c) estimated by MSCKF and VINS Mono.}
	\label{fig:calib}
\end{figure}

The 10 min long data \footnote{The data is available at \url{https://drive.google.com/open?id=1ryBXpcrpTsou-7HJDcHhq6sDI2zGCpFT}}
 for every device from the previous test were each split into 5 segments and 
processed by the two methods using the recorded focal length and frame readout time if applicable. 
The VINS Mono successfully processed all segments while the MSCKF occasionally diverged.
Ignoring the bad cases of MSCKF, the estimated lever arm, 
time delay, and frame readout time are illustrated in Fig.~\ref{fig:calib} for five devices.

The figure shows that the two methods agreed well on estimated spatial and temporal parameters. 
The large difference on the position of the camera relative to the IMU $p_{SC}$ for iPad mini is
mainly due to the fact that camera intrinsic parameters
given to VINS Mono were empirical values and are not available on iOS 9.3.

We also see that despite the tactics for syncing sensor data in the logger,
these devices were revealed to have noticeable time offsets up to 30 ms.
One notable point is that these time offset estimates had not accounted for 
the rolling shutter effect and the fact that images are timestamped at the beginning of exposure.
Subtracting half of the sum of the two effects, time offsets of iOS devices would 
look even better than those of Android devices.

\begin{figure}[]
	\centering
	\subcaptionbox{}{\includegraphics[width=0.45\columnwidth]{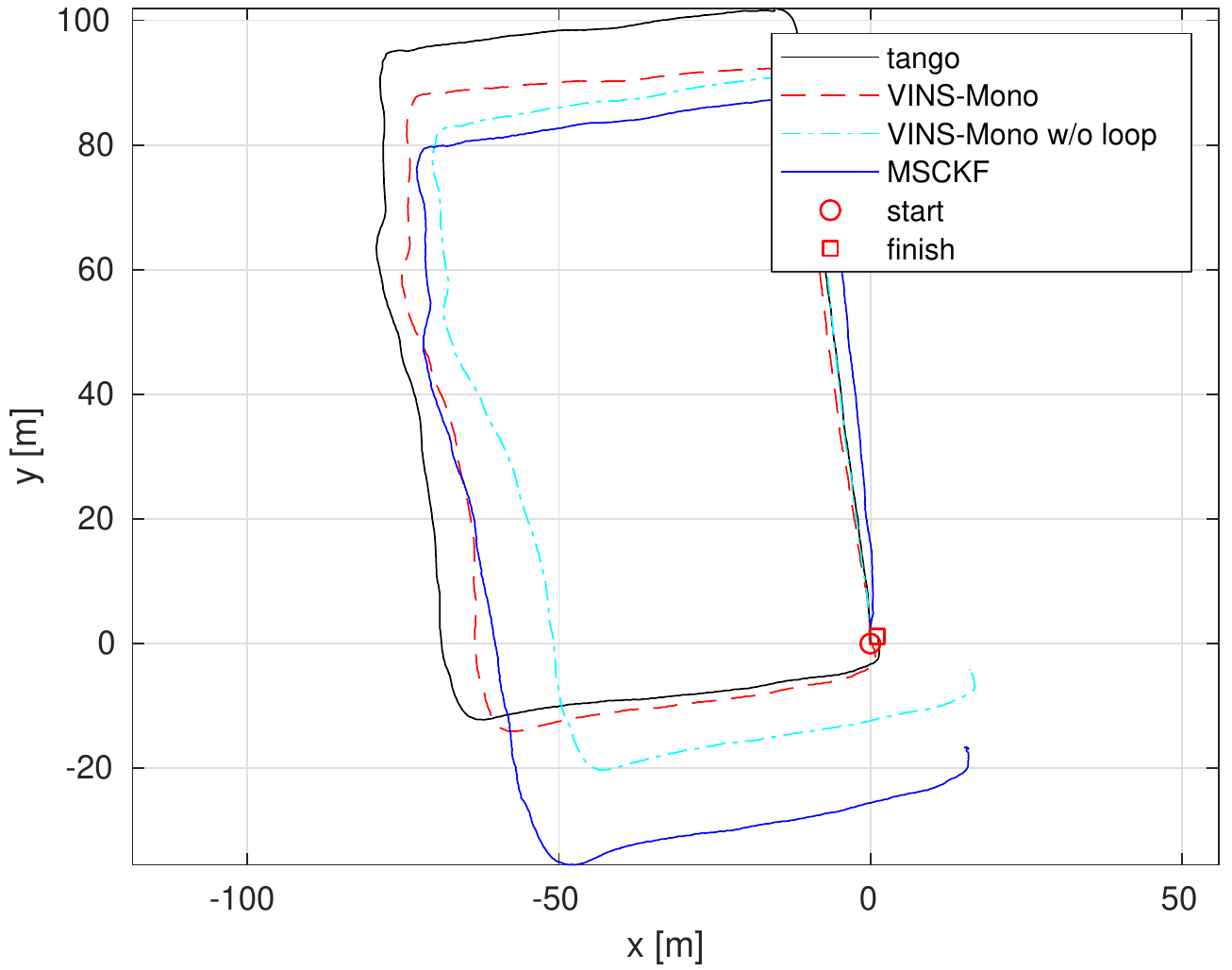}}%
	\hfill
	\subcaptionbox{}{\includegraphics[width=0.45\columnwidth]{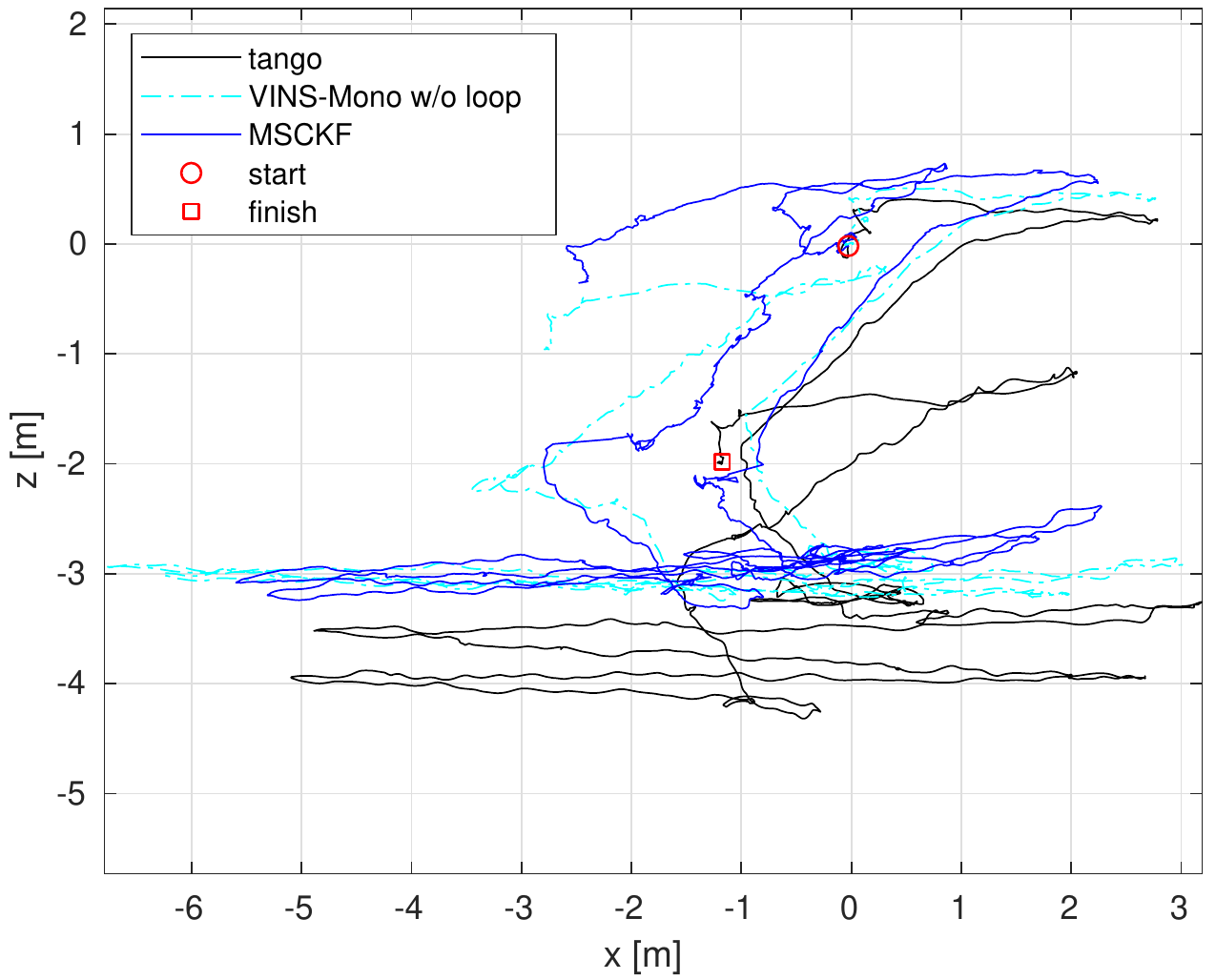}}%
	\caption{Estimated motion of a rig of a Honor View 10, an iPhone 6S, and a Lenovo Phab2 Pro compared to the Tango solution.
	a. Trajectories about a building block computed using Honor View 10 data.
	b. Traversing a stairway computed using iPhone 6S data.}
	\label{fig:traj}
\end{figure}

To show that the recorded data are amenable to SLAM methods, 
sessions of indoor and outdoor data were collected by the Honor View 10, the iPhone 6S, 
and the Tango service on the Phab2 Pro while the three devices were strapped together.
For evaluation, the start and end of a session were chosen to be the same. 
The logged data were again processed by VINS Mono and MSCKF methods.
The trajectories estimated by these methods at a stairway and a building block were 
drawn in Fig.~\ref{fig:traj} in contrast to the Tango results. 
These drawings show that the quality of the logged data is sufficient for large-scale SLAM applications.

The above tests show that the data recorded with commodity mobile devices can achieve reasonably accurate timestamps,
and they prove to be useful for SLAM and AR applications in that it is fairly easy 
to estimate sensor parameters along with the device motion with such data.

\section{Conclusion}
\label{sec:conclusion}

The mobile AR sensor (MARS) logger is presented for collecting synced visual and inertial data with commodity mobile devices at frequencies above 25Hz.
In data acquisition, the sensor characteristics including focal length,
rolling shutter skew, exposure duration are also recorded to meet the needs of SLAM algorithms.
Two versions of the logger are developed for both Android and iOS.
The quality of the datasets were investigated by two state-of-the-art SLAM algorithms,
showing that the captured datasets with devices of medium prices are suitable for mapping and localization.

We believe that along with the proliferation of mobile devices, the MARS logger will reduce the difficulty 
in capturing realistic and demanding datasets that will boost the development of the SLAM and AR algorithms.


\section*{Acknowledgment}

We thank Huan Chang for letting us participate in the iOS developer program.

\IEEEtriggeratref{4}
\IEEEtriggercmd{\enlargethispage{-0.0in}}


%
%
%

\bibliographystyle{IEEEtran}
\bibliography{IEEEabrv,ms}

\end{document}